\documentclass[10pt,conference]{IEEEtran}
\usepackage{cite}
\usepackage{amsmath,amssymb,amsfonts}
\usepackage{algorithmic}
\usepackage{graphicx}
\usepackage{textcomp}
\usepackage{xcolor}
\usepackage[skins]{tcolorbox}
\usepackage{float}
\newtcolorbox{custombox}[1]{
	colback=gray!20,
	colframe=gray!50,
	left=1mm,
	right=1mm,
	top=1mm,
	bottom=1mm,
	fonttitle=\bfseries,
	arc=2mm,
	leftrule=0mm,
	rightrule=.5mm,
	toprule=0mm,
	bottomrule=.5mm,
	notitle,
	before=\par\smallskip\noindent,
	before upper={\textbf{#1: } },
}

\begin{document}

\title{Automated Repair of Cyber-Physical Systems}

\author{\IEEEauthorblockN{Pablo Valle}
\IEEEauthorblockA{Mondragon University\\
Mondragon, Spain\\
pvalle@mondragon.edu}}

\maketitle

\begin{abstract}
Cyber-Physical Systems (CPS) integrate digital technologies with physical processes and are common in different domains and industries, such as robotic systems, autonomous vehicles or satellites. Debugging and verification of CPS software consumes much of the development budget as it is often purely manual. To speed up this process, Automated Program Repair (APR) has been targeted for a long time. Although there have been advances in software APR and CPS verification techniques, research specifically on APR for CPSs is limited. This Ph.D. research project aims to develop scalable APR techniques for CPSs, addressing problems of fault localization, long test execution times, and fitness function limitations. A new method combining spectrum-based fault localization (SBFL) with patch generation and advanced artificial intelligence techniques will be investigated. 
The approach will be validated by empirical studies on open and industrial code bases of CPSs.
\end{abstract}


\section{Motivation}
Cyber-Physical Systems (CPS) integrate digital technologies with physical processes \cite{derler2011modeling}, and can be found in different domains and industries. Debugging, testing and verification are estimated to account for 50\% to 75\% of the total software development cost budget \cite{o2017debugging}. Software debugging and verification is important as bugs are part of human nature. Because of this, Automated Program Repair (APR) has been deemed as a long-standing dream among software engineers. APR aims at providing an appropriate patch to a software program that contains a fault (i.e., a bug). Although this is a complex task, its applicability would save hundreds of hours to individual software engineers every year. For this reason, in recent years, the software engineering research community has been devising novel approaches targeting APR \cite{programr0:online}. Despite this growth, as well as the growth of articles in the field of CPS verification and testing \cite{arrieta2017search, menghi2020approximation, menghi2021theodore}, only three studies address APR of CPSs \cite{valle2023automated, jung2021swarmbug, abdessalem2020automated}. Of these three studies, two of them \cite{valle2023automated, jung2021swarmbug}, focus on misconfiguration repair. This is because CPSs are highly configurable and there may be cases where misconfiguration leads to failure of the running system \cite{valle2023automated}. The other paper \cite{abdessalem2020automated} focuses on faults in attribute interactions in autonomous driving systems. 

However, none of these studies has considered faults introduced by CPS programmers at the code level, which unlike simple software, CPS code is tightly coupled with physical components and processes. This code must interact with hardware, manage real-time constraints, and ensure safety-critical operations, making it more complex than traditional software. Additionally, CPS code must handle uncertainties inherent in physical processes, which makes faults in CPS code have more severe consequences, requiring specialized techniques for repair compared to simple software.

Traditional APR techniques can be classified into three types \cite{le2019automated}: (1) search-based, (2) constraint-based and (3) learning-based, in which recently there has been a strong trend to include Large Language Models (LLM) to provide solutions to the detected faults. Constraint-based techniques face scalability problems \cite{le2019automated} and, therefore, are not applicable in CPSs. Traditional learning techniques based on Neural Machine Translation (NMT) learn correctness patterns from different projects, limiting solutions to common bugs. This Ph.D. research work focuses on researching about the integration between the first and third type techniques.
That is, a search-based technique will be used as the main part of the solution, as it can explore multiple potential parches for bugs at the same time and is better suited for CPSs due to its ability to directly optimize within a well-defined search space, providing more predictable and verifiable solutions. Moreover, we propose to integrate the archive-based search algorithm with an LLM that helps to provide
possible fixes for detected bugs, thus overcoming some limitations of previously implemented fixes, e.g., related to fixes based on correction patterns implemented programmatically.

\section{Problem Statement and Hypothesis}
In APR, GenProg \cite{le2011genprog} the most known tool, which uses genetic programming to evolve real patches by maintaining and modifying program variants across generations. Genetic operators like crossover and mutation are applied to evolve potential solutions. Moreover, most APR techniques \cite{le2019automated} incorporate a pre-processing phase using fault localization methods such as Spectrum-Based Fault Localization (SBFL) to prioritize suspicious code fragments, improving repair efficiency. In this way, potentially faulty code elements are prioritized and modifications are targeted to the most suspicious code fragments. Inspired by GenProg, other search-based approaches, such as Arja \cite{yuan2018arja} have also been proposed. Recently, LLM-based approaches \cite{xia2023automated} have gained attention for generating large numbers of patches to fix errors. However, all these approaches face scalability challenges when applying to CPSs due to the following issues:


\textbf{I1 - Ineffective fault localization:} The number of test cases in CPSs is usually small, but test runs are long and system-wide, resulting in large code coverage with few test cases \cite{valle2022towards}. This makes current fault-finding techniques inaccurate, increasing the search space for potential patches and the likelihood of over-fitted patches.

\textbf{I2 - Long test execution time:} CPS tests are expensive and time consuming \cite{menghi2020approximation}, lasting several minutes \cite{valle2023automated, abdessalem2020automated} compared to the milliseconds of unit tests of current APR approaches \cite{le2019automated, yuan2018arja, xia2023automated}, which makes them inapplicable in the CPS domain due to the significant time required to validate potential patches. For example, initializing GenProg would require 10 hours applied to an industrial case study to which this thesis will have access \cite{valle2023automated}.

\textbf{I3 - Use of the test suite as fitness function:} Current approaches \cite{le2019automated,yuan2018arja, xia2023automated} use the number of failed and passed test cases as a fitness function, which is problematic in CPS due to the low number of test cases and the lack of internal information of test runs, deriving in almost random exploration of patches.


This Ph.D. research aims at generating new knowledge in the field of APR by proposing a novel technique that addresses these three critical issues overlooked by current methods and that to the best of my knowledge, it has not been explored so far. The technique begins with a time-aware SBFL approach to localize faults, tackling the first issue (I1). The results feed into a search algorithm that integrates LLMs with an archive-based strategy to track partial repairs, addressing the second issue (I2) and differing from existing population-based algorithms. This archive evolves new patches using two strategies: one leveraging test case details like failure criticality and timing~\cite{abdessalem2020automated}, and another utilizing CPS failure diagnosis \cite{boufaied2023trace} to assist LLMs in patch generation. This offers higher granularity in evaluating partial patches, resolving the third issue (I3) for scalability in CPSs. Based on these conjectures, the initial hypothesis of this thesis can be formulated as follows:

\begin{custombox}{Starting Hypothesis}
The inclusion of the notion of time in spectrum-based fault localization (SBFL) techniques combined with archive-based search algorithms that leverage LLMs can automate the repair process of CPSs’ software faults within an affordable time budget, saving significant debugging cost to CPS developers. 
\end{custombox}

\section{Research Objectives}

The main objective of this thesis is to research and develop a scalable technique to automatically repair software bugs in CPSs. This ambitious 
will be targeted by means of the following three tangible specific objectives:

\textbf{O1 - To increase the accuracy of fault localization techniques for the context of CPSs:} The purpose of fault localization is to identify program components with a high probability of containing an error. To achieve this, advanced techniques such as SBFL \cite{wong2016survey} will be employed. These techniques assign a suspiciousness score to code elements, (e.g., code statements) where a higher score indicates a higher probability to be faulty. Previous research \cite{liu2020efficiency} has evidenced that effective fault localization leads to (1) faster repair and (2) a lower probability of generating overfitted patches. To this end, this Objective 1 of this thesis is to improve the accuracy of fault localization techniques (1) by considering the temporal dimension in fault identification \cite{abdessalem2020automated} and (2) by tracking code elements with specific CPS features. 

\textbf{O2 – To develop a scalable automated program repair technique for the context of CPSs:} APR aims to automatically rectify faulty code, thus substantially reducing debugging costs. It takes as input a faulty program, test cases, and suspicious component classifications derived from previous objective, and outputs a set of viable patches to fix the program. 
Although numerous recent APR approaches have emerged, they lack scalability for CPSs. To solve this, a scalable solution by employing an archive-based search algorithm that uses LLMs as mutators to provide possible solutions to the detected bugs is proposed. These LLMs, together, use internal test execution data and fault diagnostics to iteratively build patches to fix the CPS bugs. Within this algorithm, mechanisms are incorporated to improve the guidance of the search~\cite{yuan2020toward, abdessalem2020automated} and avoid over-fitting of patches. In addition, this algorithm includes a patch validation mechanism due to the inherent threats of the LLMs \cite{liu2024your} (e.g., hallucination). Finally, a unified debugging approach will be explored for bugs that prove difficult to repair using automatic repair tools. 

\textbf{O3 – To validate the automated program repair technique with realistic industrial and open-source case studies:} All methodologies will be validated using appropriate empirical techniques. Practical open source case studies will be used to facilitate replication and verification of the approach within the community. In addition, industrial case studies from collaborators of the PhD student's research group (e.g., Orona, Sener, Alerion) will also be incorporated. To evaluate the developed techniques, they will be compared to state-of-the-art techniques using established metrics such as EXAM \cite{wong2016survey}, execution time or efficiency for APR techniques.

\section{Contribution and Impact}
Based on the above objectives, I expect to make the following core contributions throughout this Ph.D. research work:
    
\begin{itemize}
    \item[\textbf{C1:}]A time-aware spectrum-based fault localization approach. 
    \item[\textbf{C2:}]An open source search algorithm that combines LLMs with an archive-based strategy to keep track of partially repaired solutions.
    \item[\textbf{C3:}]A novel patch validation mechanism to ensure the correct behavior of the LLM.
    \item[\textbf{C4:}]An extension of the unified debugging approach for CPSs.
    \item[\textbf{C5:}]An open-source benchmark of case studies and a dataset of real faults of CPSs for experimentation purposes of verification techniques of CPSs.
    
\end{itemize}
Regarding the impact of this Ph.D. research work, is expected to be both national and international broad scientific impact by publishing the results in different conferences and journals. In fact, according to the APR website \cite{programr0:online} in the last few years, APR techniques have gained important attention in renowned software engineering conferences, such as ICSE, which is considered the most reputed software engineering conference. 

\section*{Acknowledgments}
Pablo Valle is part of the Software and Systems Engineering research group of Mondragon Unibertsitatea (IT1519-22), supported by the Department of Education, Universities and Research of the Basque Country. Pablo Valle is supported by the Pre-doctoral Program for the Formation of Non-Doctoral Research Staff of the Education Department of the Basque Government (Grant n. PRE\_2024\_1\_0014).

\bibliographystyle{IEEEtran}
\bibliography{references}

\end{document}